\documentstyle[sprocl]{article}
\input{psfig}
\begin{document}

{\hfill LBL-40457}
\vskip .01 in
{\hfill June, 1997}

\title{The Gold Flashlight: Coherent Photons (and Pomerons) at RHIC}

\author{Spencer Klein and Evan Scannapieco}

\address{Lawrence Berkeley National Laboratory, Berkeley, CA, 94720, USA}

\def\gamgam{$\gamma\gamma\,$}
\def\gampom{$\gamma P$}
\def\pompom{$PP$}
 
\maketitle\abstracts{The Relativistic Heavy Ion Collider (RHIC) will
be the first heavy ion accelerator energetic enough to produce
hadronic final states via coherent \gamgam, \gampom, and \pompom\
interactions.  Because the photon flux scales as $Z^2$, up to an
energy of about $\gamma\hbar c/R\approx 3$ GeV/c, the \gamgam\
interaction rates are large.  RHIC $\gamma P$ interactions test how
Pomerons couple to nuclei and measure how different vector mesons,
including the $J/\psi$, interact with nuclear matter.  $PP$ collisions
can probe Pomeron couplings.  Because these collisions can involve
identical initial states, for identical final states, the \gamgam,
\gampom, and \pompom\ channels may interfere, producing new effects.
We review the physics of these interactions and discuss how these
signals can be detected experimentally, in the context of the STAR
detector.  Signals can be separated from backgrounds by using
isolation cuts (rapidity gaps) and $p_\perp$.  We present Monte Carlo
studies of different backgrounds, showing that representative signals
can be extracted with good rates and signal to noise ratios.}

\centerline{Presented at {\it Photon '97}, May 10-15, 1997, Egmond aan Zee,
The Netherlands}

\section{Physics Processes}
The Relativistic Heavy Ion Collider\cite{RHICLUM} 
(RHIC) will be energetic enough to produce
massive final states via  
\gamgam, \gampom, and \pompom\ interactions that coherently couple
to the nuclei as a whole.
As the number of virtual photons associated with each nuclei goes as 
$Z^2$ up to a photon energy of 
approximately  
$\gamma\hbar c/R\approx 3$ GeV/c, the \gamgam  rate at intermediate
energies will be comparable to those of the next generation $e^+ e^-$
colliders.   RHIC will also produce 
a high number of coherent photon-Pomeron
interactions (\gampom) and 
two-Pomeron interactions.

\subsection{\gamgam Interactions}
The luminosity of
\gamgam\ collisions at heavy ion colliders has been discussed by
several authors\rlap.\cite{candj} \cite{baurandff} \cite{hencken}
To avoid events where hadronic particle production
overshadows the \gamgam\ interaction, events where the nuclei
physically collide (with impact parameter $b<2R_A$, $R_A$ being the
nuclear radius) are excluded from calculations of the usable
luminosity.  This reduces the luminosity by about 50\%,
depending on the energy.  

The usable \gamgam\ luminosity for gold, copper and iodine collisions at RHIC
design luminosity is given in the left panel of
Fig.~1.  The lighter nuclei benefit from
the higher $AA$ luminosity, slightly higher beam energy, and smaller
nuclear radius, which more than compensates for the reduced $Z$.
Comparison curves for CLEO at CESR and LEP2 are also shown.

\begin{figure} 
\centerline{
\psfig{figure=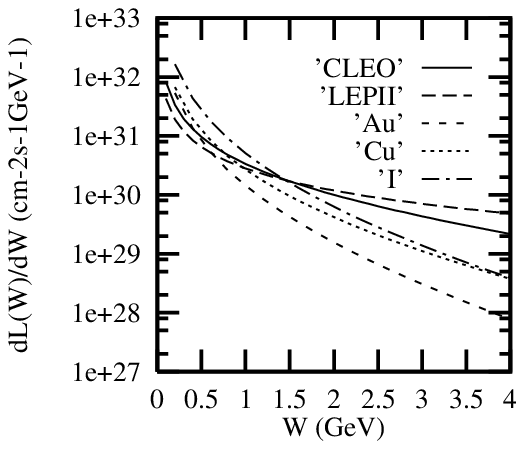,rheight=1.7 in,width=2.4 in} 
\psfig{figure=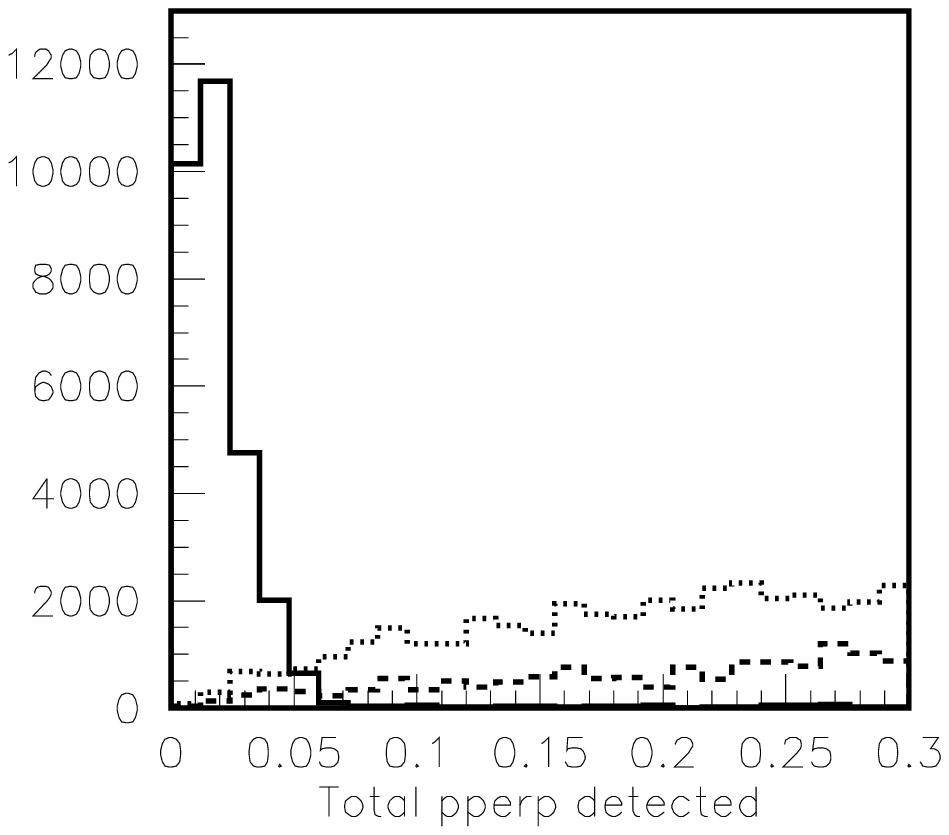,rheight=1.7 in,width=2.4 in}}
\caption{Left: Comparison of \gamgam\ luminosities at RHIC, for gold, iodine
and copper beams,  with those of
CESR(CLEO) and LEP II.  Right:
Comparison of $p_\perp$ between \gamgam and background passing our cuts. 
The solid
curve is for $\rho^0 \rho^0$ production near threshold.  The short
dashes are beam-gas and the long dashes are peripheral nuclear backgrounds.}
\label{fig:luminosity}
\end{figure}

Due to the nuclear form factor, the
photons are almost real, with a $Q^2$ cutoff given by the nuclear
size, about (30 MeV/c)$^2$ for gold.  Because of this
cutoff, the perpendicular momentum of the photons
is small, $p_\perp < \hbar c/R$; this is important for
separating coherent from incoherent interactions.  This is illustrated
in the right panel of Fig 1.

\subsection{\gampom Interactions}

\gampom\ interactions on proton targets have been studied
extensively at HERA.  RHIC can study these interactions in a nuclear
environment.  For the reaction $\gamma P\rightarrow V$, where $V$ is a
vector meson, RHIC will reach higher center of mass energies and
luminosities than the NMC \cite{NMC} and E-665 \cite{E665} studies,
producing 100,000's of exclusive $\rho$ and $\phi$ mesons per year,
along with large numbers of excited states.  RHIC will also produce
significant numbers of $J/\psi$.  In the Vector Dominance Model, these
rates measure the interaction between the vector meson and the
nucleus\rlap.\cite{brodsky} Measurements of how vector meson production
scales with $A$ can probe meson absorption by nuclear matter.  Because
meson scattering has a similar form factor to the photon coupling,
this reaction has similar kinematics to \gamgam\ processes.

\subsection{\pompom \ Interactions and Interference Measurements}

Unobscured \pompom\ interactions can only occur in the impact
parameter range $2R_A+2R_P > b > 2R_A$, where $R_P$ is the range of
the Pomeron.  A measurement of the \pompom\ cross section can thus
measure the range of the Pomeron.  The difficulty in this measurement
is separating \gamgam\ and \pompom\ interactions; the two reactions
have very similar kinematics and a statistical separation is required.
However the relative rates will change as $A$ varies; for protons,
\pompom\ interactions will dominate, while \gamgam should dominate for
Au.  The \gamgam\ luminosity can be measured from
$\gamma\gamma\rightarrow e^+e^-$ and the \pompom\ luminosity found by
subtraction.  It may also be possible to use impact parameter
dependent signals of nuclear breakup to better distinguish $\gamma$
and $P$ emission\rlap.\cite{hencken}

The similarity between \gamgam, \gampom\ and \pompom\ events allows
for the possibility of interference between the two channels.  One
example is dilepton production from $\gamma\gamma\rightarrow e^+e^-$
and $\gamma P\rightarrow V\rightarrow e^+e^-$; the two channels can
interfere, and a measurement of the phase of the interference is
sensitive to the real part of the Pomeron and the interaction of the
vector meson with the nuclear potential\rlap.\cite{leith}

\section{Experimental Feasibility}
For any of these measurements to be feasible, it must be possible to
separate these coherent interactions from incoherent backgrounds at
both the trigger and analysis levels\rlap.\cite{photon95}  The major
backgrounds that we have identified are grazing nuclear collisions,
photo-nuclear interactions, beam gas interactions, debris from
upstream beam breakup, and cosmic ray muons; the latter two only
affect triggering.

Two useful factors for separating these signals from backgrounds are
rapidity gaps and perpendicular momentum.  We have concentrated on
final states that can be completely reconstructed.  We then require
that the detector contain nothing except the final state in question.
For central events, this naturally reduces to requiring rapidity gaps.
Because of the coherence, the $p_\perp$ scale is $\sqrt2\hbar c/R$,
about 45 MeV/c for gold, much smaller than the typical hadronic
momentum scale of 300 MeV/c.

\subsection{STAR}

The Solenoidal Tracker at RHIC (STAR) is a general purpose large
acceptance detector\rlap.\cite{STAR} A time projection chamber tracks
charged particles with pseudorapidity $-2<\eta<2$.  A silicon vertex
tracker measures impact parameter over $-1<\eta<1$.  A time of flight
(TOF) system and $dE/dx$ in the TPC help with particle
identification. Two forward TPCs are sensitive to charged particles
with $2.5 < | \eta | < 4 $, and an electromagnetic calorimeter detects
photons in the range $-1<\eta<2$.

STAR has a multi-level trigger which is well suited to studying
peripheral collisions.  Scintillators and wire chamber readouts
surrounding the TPC measure the charged multiplicity for $-2<\eta<2$
on each beam crossing.  Events are selected based on multiplicity and
topology.  At higher trigger levels, the calorimeter can contribute to
the trigger and TPC tracking information can be used to select events
based on the location of the event vertex and total $p_\perp$.

\subsection{Signal and Background Simulation}

We have performed Monte Carlo calculations of the \gamgam\ signals and
backgrounds from grazing nuclear and beam gas
interactions\rlap.\cite{starnote} Other backgrounds have been
estimated by scaling and other methods.

We calculated tables of \gamgam\ luminosity as a function of invariant
mass and rapidity, and then generated simulated events based on these
tables.  Transverse momentum spectra were included using a Gaussian
form factor with a characteristic width of $1/R$.  Cuts were applied
to simulate the detector acceptance and planned analysis procedure.

Grazing nuclear collisions and beam gas events were simulated using
both the FRITIOF and Venus nuclear Monte Carlos.  These events were
subject to the same cuts.  Photo-nuclear collision rates were
estimated by scaling from the beam gas rates, making use of the
similar kinematics; a more detailed estimate is in progress.

To determine the feasibility of studying \gamgam\ interactions with
STAR, we have considered 3 sample analyses: $\gamma \gamma \rightarrow
f_2(1270) \rightarrow \pi^+ \pi^-$, $ \gamma \gamma \rightarrow \rho^0
\rho^0 \rightarrow \pi^+ \pi^- \pi^+ \pi^-$ and
$\gamma\gamma\rightarrow\eta_c\rightarrow K^{*0}K^-\pi^+$.  These
reactions were chosen to be representative of a wide range of
reactions that produce two or four charged particles in the STAR TPC.
To separate these events from backgrounds, we have applied cuts to the
charged and neutral multiplicity visible in STAR, required that
$p_\perp < 100$ MeV, and required an appropriate invariant mass cut.
The predicted rates and backgrounds for these analyses are given in
Table \ref{tab:rates}.  Although the FRITIOF and Venus predictions are
very different, this analysis shows that $f_2(1270) \rightarrow \pi^+
\pi^-$ and $ \gamma \gamma \rightarrow \rho^0 \rho^0 \rightarrow \pi^+
\pi^- \pi^+ \pi^-$ reactions should be clearly separable from
backgrounds, while more challenging measurements such
$\Gamma(\gamma\gamma)$ for the 2960 MeV $J^{PC}=0^{-+}$ $c \overline
c$ resonance $\eta_c \rightarrow K^{*0}K^-\pi^+$ may be possible with
appropriate particle identification by TOF and $dE/dx$.

\begin{table}[t]
\caption{Rates and backgrounds for \gamgam events for gold on gold
collisions at RHIC for 3 sample analyses.  The $\rho^0\rho^0$ events
were near threshold, with invariant masses between 1.5 and 1.6
GeV/c$^2$. The last line assumes particle identification by $dE/dx$ and
TOF.\hfill}
\label{tab:rates}
\vspace{0.1cm}
\begin{center}
\begin{tabular}{|l|r|r|r|r|}
\hline
Channel & Efficiency & Detected & FRITIOF & Venus \\
 & & Events/Yr & Background & Background \\
\hline
$f_2(1270) \rightarrow \pi^+ \pi^-$ & 85\% & 
$9.2 \times 10^5$  & 53,000 & 100,000\\
$\rho^0 \rho^0 \rightarrow  \pi^+ \pi^- \pi^+ \pi^-$ & 38\% &  
$1.6 \times 10^4$ & 3,500 & 1,400 \\
$\eta_c \rightarrow K^{*0}K^-\pi^+$ & 57\% & 
70 &  210 & 510 \\ 
$\eta_c$ (w/ PID) & 57\% & 
70 & 8 & 20 \\ 
\hline
\end{tabular}
\end{center}
\end{table}

We have also considered the problem of triggering on these events.  In
addition to the grazing nuclear collisions, beam gas events and
photonuclear interactions, at the trigger level there are backgrounds
from beam nuclei interactions upstream of the detector and cosmic ray
muons.  Monte Carlo studies have shown that, using the multi-level
trigger in STAR, it is possible to devise trigger algorithms with good
acceptance for coherent interactions and good background rejection.
The trigger algorithms are based on requiring two or four tracks in
the central TPC, with nothing else visible in the detector.  
\vskip .05 in
\noindent
{We would like to thank our colleagues in the STAR collaboration
for their advice and support.  This work was supported by 
the U.S. DOE under contract DE-AC-03-76SF00098.  E.S. was partially
supported by an NSF Fellowship.
\vskip 0.1 in}
{\bf References \rm}
\def\etal{\it et al.\rm}

\end{document}